\documentclass[aps,prl,reprint,superscriptaddress]{revtex4-2}
\usepackage[utf8]{inputenc}
\usepackage{amsthm}
\usepackage{amsmath}
\usepackage{amssymb}
\usepackage{physics}
\usepackage{color}
\usepackage{hyperref}
\usepackage{graphicx,xcolor}
\usepackage{caption}
\usepackage{subcaption}
\usepackage{overpic}

\newcommand{\mean}[1]{\mathbb{E}\left[#1\right]}

\newcommand{\cavg}[1]{\langle#1\rangle} 

\begin{document}
\title{Genuine Quantum effects in Dicke-type Models at large atom numbers}

\author{Kai Müller}
\affiliation{Institut für Theoretische Physik, Technische Universität Dresden, D-01062 Dresden, Germany.}
\author{Walter T. Strunz}
\affiliation{Institut für Theoretische Physik, Technische Universität Dresden, D-01062 Dresden, Germany.}

\begin{abstract}
    We investigate the occurrence of genuine quantum effects and beyond mean-field physics in the balanced and unbalanced open Dicke model with a large, yet finite number of atoms $N$. Such driven and dissipative quantum many-body systems have recently been realized in experiments involving ultracold gases inside optical cavities and are known to obey mean-field predictions in the thermodynamic limit $N\to\infty$. 
    Here we show quantum effects that survive for large but finite $N$, by employing a novel open-system dynamics method that allows us to obtain numerically exact quantum dynamical results for atom numbers up to a mesoscopic $N\approx 1000$. 
    While we find that beyond-mean-field effects vanish quickly with increasing $N$ in the balanced Dicke model, we are able to identify parameter regimes in the unbalanced Dicke model that allow genuine quantum effects to persist even for mesoscopic $N$. They manifest themselves in a strong squeezing of the steady state and a modification of the steady-state phase diagram that cannot be seen in a mean-field description. This is due to the fact that the steady-state limit $t\rightarrow \infty$ and thermodynamic limit $N\rightarrow \infty$ do not commute.
\end{abstract}
\maketitle

\paragraph{Introduction.---}
Recent years have wittnessed considerable experimental progress in realizing driven and dissipative quantum many-body systems across different fields \cite{reviewCQED_Francesco, reviewDDQMBS_UA}. 
These developments have been stimulated by significant theoretical efforts to both understand and model these systems, as well as by potential applications \cite{review_disspation_engineering, quantum_battery, associativeMemoryPRX,associative_memory_Lesanovsky}. 
In particular, experiments with ultracold quantum gases in optical cavities allow for a precise preparation, manipulation and control of non-thermal equilibrium steady states in combination with large light-matter coupling strengths. 
Following early pioneering works \cite{PhysRevA.67.051802,PhysRevLett.91.203001,PhysRevLett.99.213601,Colombe2007Nov,Brennecke2008Oct}, these platforms have recently been considered for quantum optimization schemes \cite{NQueensSolver,annealing}, quantum sensing \cite{gravimeter_cqed} as well as for the controlled study of (driven-dissipative) quantum many-body Hamiltonians \cite{chemistrySimulation,tunable_range_interaction}.
One prominent example of the latter is the (Hepp-Lieb) Dicke Model \cite{HeppLieb, Introduction_Dicke}, 
which can nowadays be implemented using momentum states of a BEC \cite{Dicke_BEC_theory, Baumann2010Apr,Kessler2014May, Kollar2015Apr} or a 4-level scheme based on stimulated Raman transitions \cite{Dicke_4lvl_theory, Dicke_4lvl_experiment} in experiments involving $N\approx 10^4\,-\,10^5$ atoms.
The model has attracted great interest as it features a phase transition
from a normal to a superradiant state, regular as well as chaotic dynamics 
and, in addition, serves as a starting point for many other models based on light-matter interaction \cite{Stitely_dynamics_TC}.
More recently, generalized or unbalanced Dicke models have been investigated in theory \cite{gen_Dicke_semiclassic, gen_Dicke_phases_Zilberberg, gen_Dicke_finite_size,gen_Dicke_finite_T,gen_Dicke_Zilberberg_diss_stab_norm_phase} and experiment \cite{gen_Dicke_Singapur, genDicke_ETH}. 
There, an even richer phase diagram emerges from the strong light-matter coupling, resulting in limit cycles and bistability regions in addition to the normal and superradiant phases. 
However, in single mode optical cavities the effective infinite-range interactions mediated by the light field constrain the physics in the thermodynamic limit to mean-field effects \cite{Lesonvsky_meanfield_exact}, making it an outstanding challenge to search for and observe genuine quantum phenomena.
While our understanding of the dynamics and phase transitions in these driven, dissipative quantum many-body systems has increased tremendously, the question of true quantum advantages of these setups over classical realizations \cite{classical_Dicke_Experiment} is still unanswered.
While the exact diagonalization of the model for a small number of atoms $N$ \cite{DickeSuperradianceKirtonKeeling,gen_Dicke_finite_size,gen_Dicke_coin_toss} as well as mean-field equations for the thermodynamic limit $N\to\infty$ \cite{gen_Dicke_semiclassic,gen_Dicke_semiclassics_Keeling,limit_cycles_mf} have been studied thoroughly in previous works, it remains an open question which genuine quantum effects can be found in the experimentally relevant regime of large but finite $N$. Alternatively, we may ask 
how low would $N$ need to be in order to observe and possibly use genuine quantum phenomena.
Here we tackle these questions by deploying a newly developed quantum dynamics method based on the Hierarchy of Pure States (HOPS) \cite{HOPS, HOPS_Richard}, which is able to bridge the fully quantum to the (presumably classical) mean-field regime by allowing numerically exact simulations of the quantum non-equilibrium dynamics for up to $N\approx 1000$ atoms in Dicke-type models. 
Furthermore, we achieve these results in the arguably most challenging parameter regime, where all terms in the Hamiltonian as well as the dissipation contribute on the same time scale and no simplifying assumptions can be made. 
We can thus investigate the deviations of the exact solution from the mean-field approximation as well as the occurrence of genuine quantum effects at large $N$. Most notably, we find that these finite size effects lead to a modification of the quantum phase diagram that cannot be explained by mean-field: the steady-state $t\rightarrow \infty$, and thermodynamic $N\rightarrow \infty$ limits do not commute. In contrast to mean field, where we take the latter limit before the former, our studies here determine steady states for large but finite $N$, and we only consider the $N\rightarrow \infty$ limit after determining the steady states for finite (large) $N$.
\paragraph{Generalized open Dicke model and the nuHOPS method.---}
We focus our considerations on the unbalanced open Dicke model, also known as the Interpolating-Dicke-Tavis-Cummings 
model with Hamiltonian ($\hbar = 1$)
\begin{equation}\label{eq:Dicke_H}
\begin{split}
    H =& \omega_a S^z +\frac{2\bar{g}}{\sqrt{N}} S^x(a+a^\dagger) + \frac{2i\Delta g}{\sqrt{N}} S^y(a-a^\dagger) + \omega_c a^\dagger a,\\
    =&\omega_a S^z +\frac{1}{\sqrt{N}}\left(g_-S^-a^\dagger + g_+ S^+a^\dagger +h.c. \right) + \omega_c a^\dagger a,
\end{split}
\end{equation}
Here, $a$ ($a^\dagger$) is the cavity photon annihilation (creation) operator satisfying $[a,a^\dagger] = 1$, $S^j$ are the collective spin operators obeying the commutation relation $[S^j, S^k] = i\epsilon_{jkl} S^l$ and $\bar{g} = (g_+ + g_-)/2$, $\Delta g = (g_- - g_+)/2$.
This unbalanced Dicke model arises in cavity-QED experiments 
\cite{gen_Dicke_Singapur, genDicke_ETH, gen_Dicke_semiclassic} and it reduces to the 
Hepp-Lieb Dicke model \cite{HeppLieb, Introduction_Dicke} for vanishing imbalance $\Delta g=0$ and to the Tavis-Cummings model \cite{JC_Model} for $g_+ = 0$.
The primary source of dissipation in experiments \cite{gen_Dicke_Singapur, genDicke_ETH} is the leakage of cavity photons (rate $\kappa$), modelled by the standard master equation
\begin{equation}\label{eq:openDicke}
    \dot\rho_{tot} = -i[H, \rho_{tot}] + \kappa\left(2a\rho_{tot} a^\dagger - \{a^\dagger a, \rho_{tot}\}\right).
\end{equation}
For the remainder of this letter we consider the spin degree of freedom as our ``open system'', evolving with 
Hamiltonian $H_\mathrm{sys}=\omega_a S^z$ and the damped cavity mode as its (non-Markovian) bath. Our state of interest is thus 
\begin{equation}\label{eq:openDickeState}
    \rho=\operatorname{Tr}_{cavity}(\rho_{tot}).
\end{equation}
Determining the exact solution of Eq. \eqref{eq:openDicke} at large but finite $N$ requires a powerful method capable of dealing with the large system Hilbert space and the possibly highly excited cavity mode.
For this task we employ a variant of the HOPS method \cite{HOPS, HOPS_Richard}\footnote{The HOPS algorithm is publically available online under \url{https://github.com/OpQuSyD/}\label{GitHubHOPS}}, which is a pure-state trajectory method for non-Markovian open quantum systems based on the non-Markovian quantum state diffusion framework \cite{NMQSD}.
Details of this newly developed near-unitary Hierarchy of Pure States (nuHOPS) method are published in an accompanying article \cite{PRA_placeholder} and will not be subject of the present letter.
For the unbalanced Dicke model our procedure amounts to an exact mapping of the open system dynamics of \eqref{eq:openDicke} with \eqref{eq:openDickeState} to the following nuHOPS quantum trajectory equation:
\begin{equation}\label{eq:shiftedHOPS}
    \begin{split}
        \partial_t \ket{\Phi_t} =& \Big[-i H_{sys} + ({z}_t^* + \mu^*(t))L -\mu(t) L^\dagger
        -(i\omega_c+\kappa)b^\dagger b\\
        &-2i\bar{g}/\sqrt{N} \left(\left(L^\dagger - \cavg{L^\dagger}_t\right)b + \left(L - \cavg{L}_t\right)b^\dagger\right)\Big]\ket{\Phi_t}.
    \end{split}
\end{equation}
Here, we set $L:=S^x - i\frac{\Delta g}{\bar{g}} S^y$ and $b$ ($b^\dagger$) is an auxiliary bosonic annihilation (creation) operator reflecting the hierarchy of equations in (nu-)HOPS. Furthermore, we introduce $\mu(t)=\int_0^t\alpha(t-s) \cavg{L}_s$ and choose $z_t^*$ to be an Ornstein-Uhlenbeck process with mean $\mean{z_t^*} = 0$ and correlation $\mean{z_t z_s^*} = (2\bar{g})^2 \exp{-i\omega_c(t-s) - \kappa|t-s|}$. The physically relevant quantum trajectories are the auxiliary oscillator vacuum components $|\psi_t^{(0)}(z^*)\rangle = \langle 0_b|\Phi_t \rangle$ of the nuHOPS state. These are used to determine the expectations values $\langle L \rangle_t = \langle\psi_t^{(0)}(z^*)|L|\psi_t^{(0)}(z^*)\rangle / \langle\psi_t^{(0)}(z^*)|\psi_t^{(0)}(z^*)\rangle$.
The desired density operator of the spin state \eqref{eq:openDickeState} is recovered without approximation from the ensemble average $\rho = \mean{|\psi_t^{(0)}(z^*)\rangle\langle\psi_t^{(0)}(z^*)|/\langle\psi_t^{(0)}(z^*)|\psi_t^{(0)}(z^*)\rangle}$ over the stochastic process $z_t^*$. \\
Solving Eq.~\eqref{eq:shiftedHOPS} instead of Eq.~\eqref{eq:openDicke} gives us three key numerical advantages. 
(1) Since each trajectory remains pure we can evolve state vectors instead of density matrices which would scale with the square of the Hilbert space dimension. 
(2) Through the function $\mu(t)$, Eq.\eqref{eq:shiftedHOPS} takes into account explicitly the stochastic mean-field evolution of the cavity, such that the $b$ ($b^\dagger)$ operators need only describe the fluctuations around the mean field and $\cavg{b} \approx 0$ during the dynamics. The nuHOPS state can thus be determined by a much more efficient Fock space truncation.
(3) As each trajectory remains localized we employ an adaptive basis to further reduce the dimension of the propagated state vector.
Making use of these key advantages 
allows us to solve \eqref{eq:openDicke} exactly for up to 1000 atoms, which is an improvement of one order of magnitude compared to previous results \cite{DickeSuperradianceKirtonKeeling}. 
\paragraph{Open Dicke Model.---}
First, we turn our attention to the balanced case ($\Delta g =0$, $L=S^x$).
Here, it is known that in the thermodynamic limit $N\to\infty$, the expectation values $\vec m = \cavg{\vec S}/N$ and $\beta = \cavg{a}/\sqrt{N}$ evolve exactly according to the mean-field equations~\cite{Lesonvsky_meanfield_exact} 
\begin{equation}\label{eq:meanfieldDicke}
    \begin{split}
        \dot{\vec m} =& \omega_a \vec e_z \cross \vec m + 2\bar{g} (\beta + \beta^*) \vec e_x \cross \vec{m} ,\\
        \dot \beta =& -\left(i\omega_c + \kappa\right)\beta - igm_x,
    \end{split}
\end{equation}
with unit vectors $\vec e_x$, $\vec e_z$. Solving these equations for the steady state reveals the superradiant phase transition, as one finds the normal state $\vec{m} = (0,0,-1/2)$, $\beta=0$ below a critical coupling strength $2\bar{g} < g_c = \sqrt{\omega_a\left(\omega_c^2+\kappa^2\right)/\omega_c}$, and a pair of superradiant states with $m_z>-1/2$, $\beta>0$ above the critical coupling.

\begin{figure*}[t]
    \begin{minipage}{\textwidth}
        \centering
        \begin{subfigure}[b]{0.24\textwidth}
             \centering
             \includegraphics[width=\textwidth]{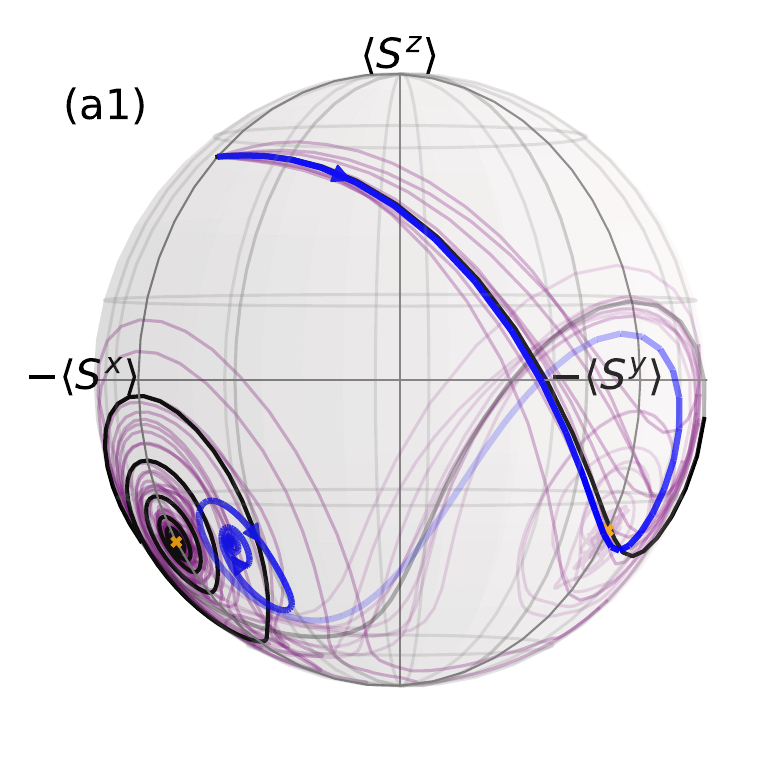}
         \end{subfigure}
         \begin{subfigure}[b]{0.24\textwidth}
             \centering
             \includegraphics[width=\textwidth]{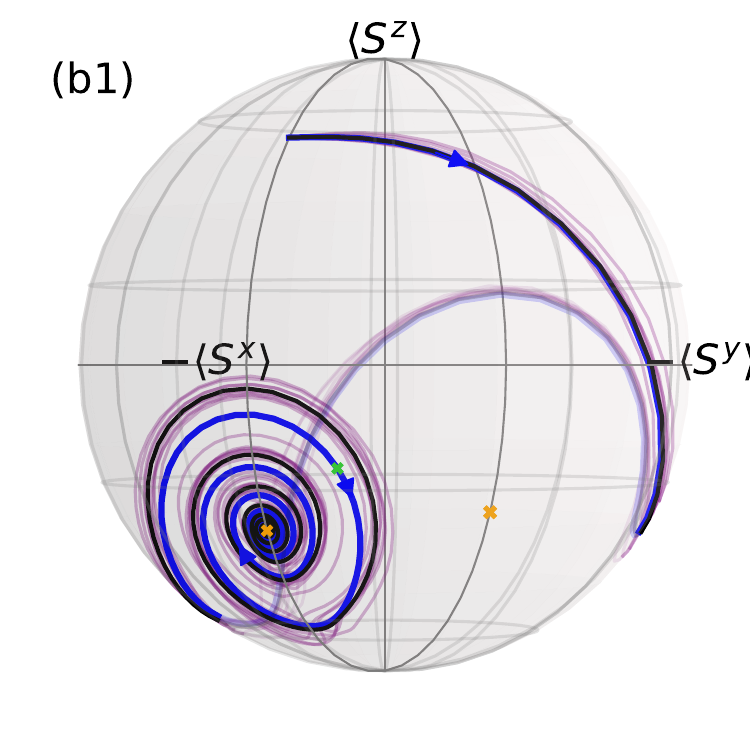}
         \end{subfigure}
         \begin{subfigure}[b]{0.24\textwidth}
             \centering
             \includegraphics[width=\textwidth]{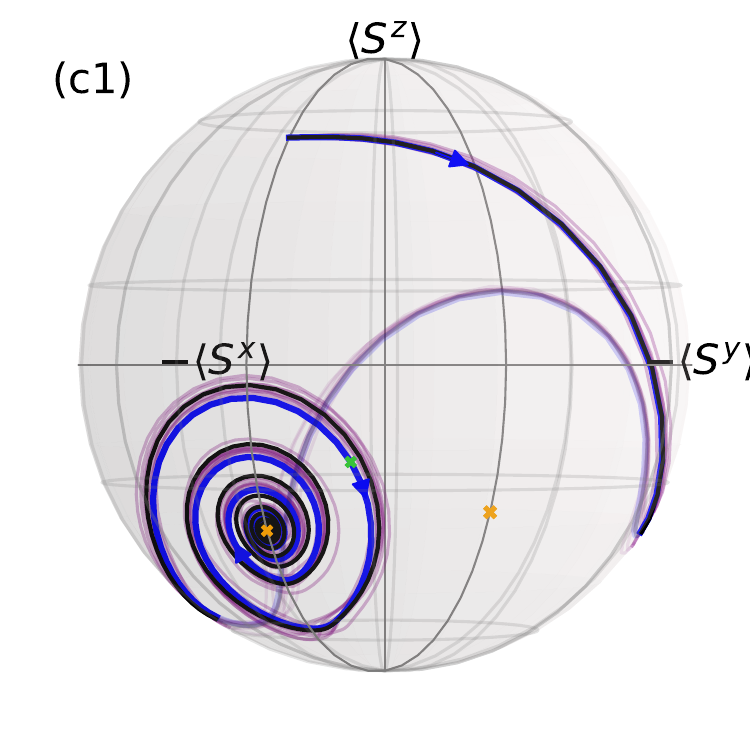}
         \end{subfigure}
         \begin{subfigure}[b]{0.24\textwidth}
             \centering
             \includegraphics[width=\textwidth]{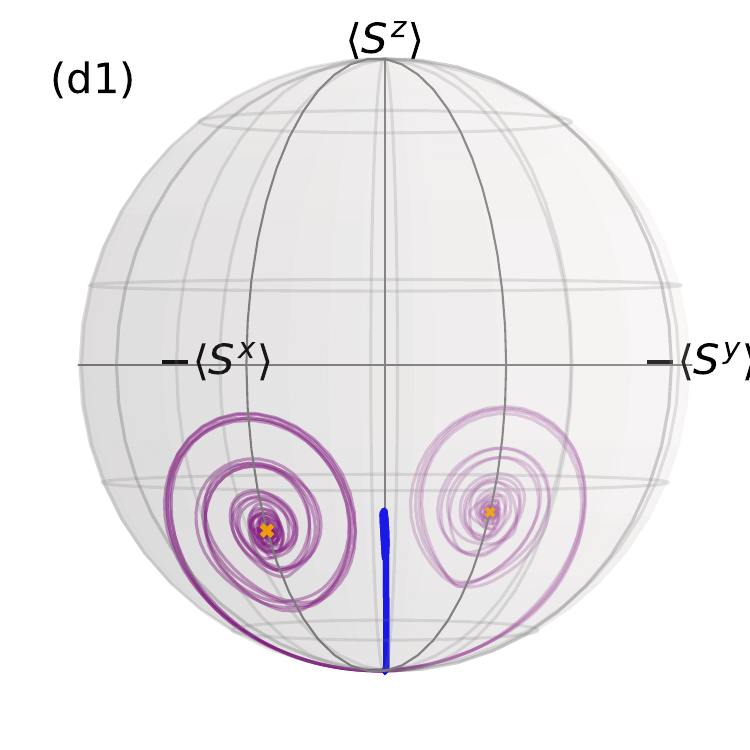}
         \end{subfigure}
         \begin{subfigure}[b]{0.24\textwidth}
             \centering
             \includegraphics[width=\textwidth]{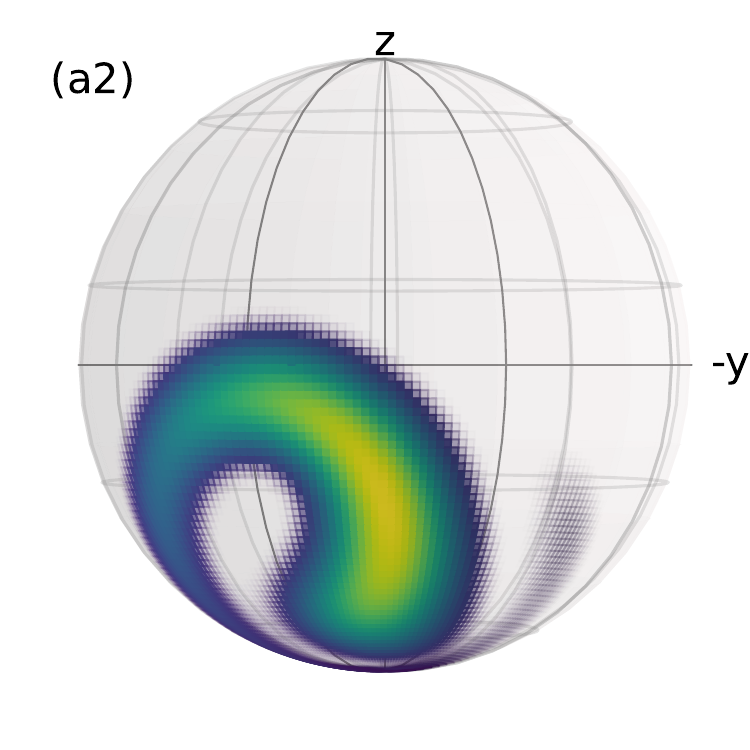}
         \end{subfigure}
         \begin{subfigure}[b]{0.24\textwidth}
             \centering
             \includegraphics[width=\textwidth]{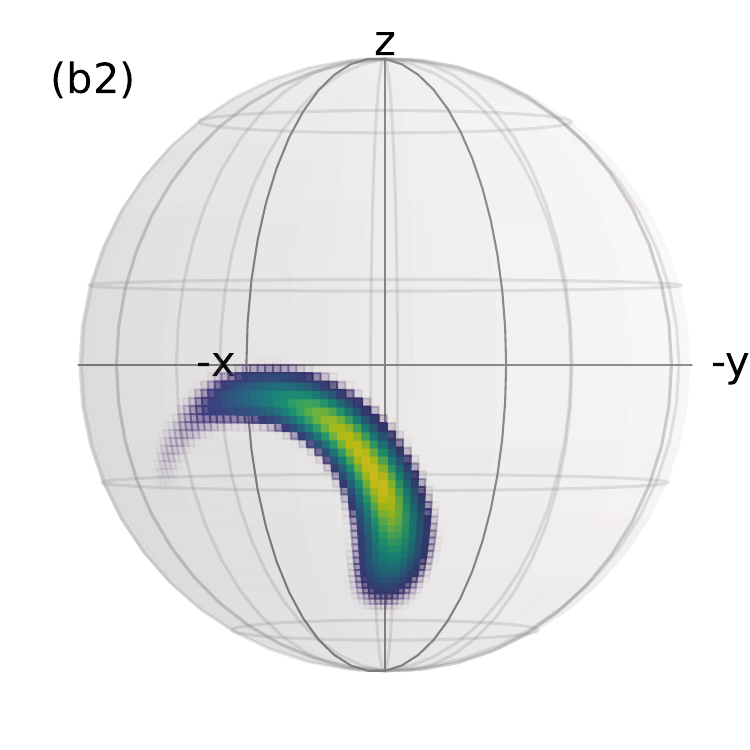}
         \end{subfigure}
         \begin{subfigure}[b]{0.24\textwidth}
             \centering
             \includegraphics[width=\textwidth]{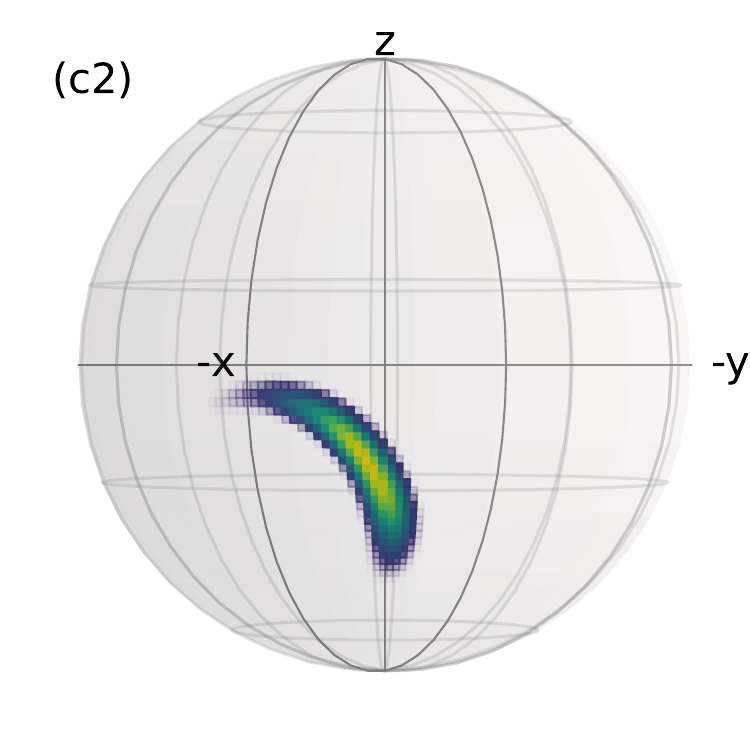}
         \end{subfigure}
         \begin{subfigure}[b]{0.24\textwidth}
             \centering
             \includegraphics[width=\textwidth]{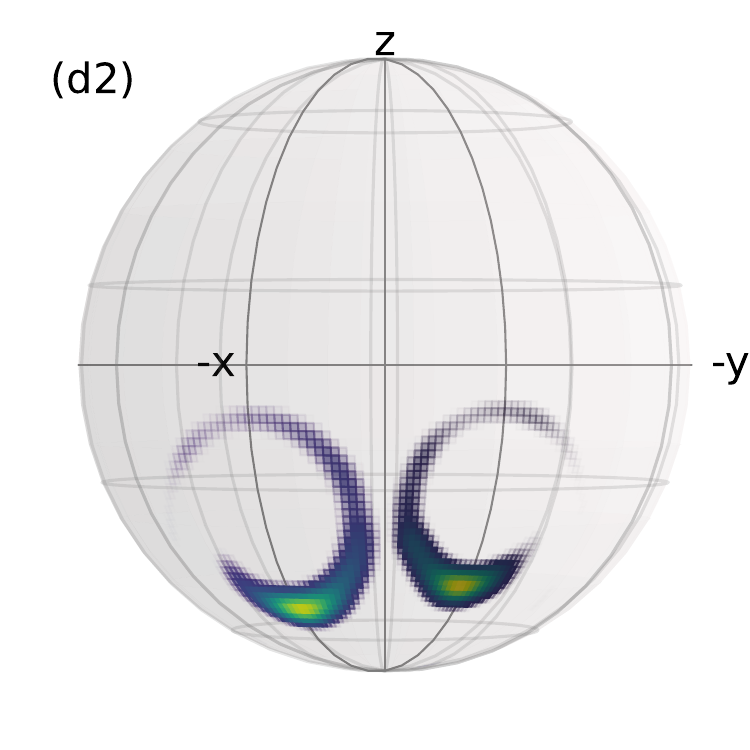}
         \end{subfigure}
    \caption{Numerically exact dynamics of the spin state in the (balanced) open Dicke model. Plots (a1) to (d1) show the quantum dynamics of the spin expectation values on a generalized Bloch sphere (thick blue line) compared to the dynamics obtained from the mean-field equations \eqref{eq:meanfieldDicke} (black line). Plots (a1) to (c1) show the dynamics for $N = 100, 500, 1000$ respectively with the initial state $\ket{\Psi (0)} = \ket{\theta=\pi/4, \phi=\pi}\otimes\ket{0}$, where $\ket{\theta, \phi}$ denotes a spin coherent state.
    (d1) shows the dynamics for the initial state $\ket{\Psi (0)} = \ket{\theta=\pi, \phi=0}\otimes\ket{0}$ and $N=1000$, corresponding to a quench from the free spin and field Hamiltonian. Selected trajectories are shown as thin purple lines and the mean-field steady-state predictions are marked by orange crosses. Curves inside or on the backside of the sphere are shown increasingly transparent. Plots (a2) - (d2) show the spin-Q function $Q(\theta, \varphi) = \bra{\theta, \varphi}\rho\ket{\theta, \varphi}$ at a fixed time $t_1 = 7.5/\omega_a$, marked by a green dot in the upper plots.}
    \label{fig:blochDynamics}
    \end{minipage}
\end{figure*}
With our nuHOPS method Eq.~\eqref{eq:shiftedHOPS}, we determine the exact numerical solution of the reduced spin state of Eq.~\eqref{eq:openDicke}. We focus on a challenging regime with $\kappa/\omega_a = 0.5 $ and $\omega_c/\omega_a = 2.5$, where typical approximation schemes, like adiabatic elimination of the cavity mode \cite{Jäger_good_adiabatic_elimination}, are not applicable. 
The parameters are chosen to resemble the experiment in Ref. \cite{Klinder2015Mar}. The spin dynamics for $g=1.4g_c$ are shown in Fig.~\ref{fig:blochDynamics} for different numbers of atoms and different initial conditions.\\
The upper plots (a1) to (d1) show the dynamics of the spin expectation values visualized on a generalized Bloch sphere (thick blue line) compared to the dynamics obtained from the mean-field equations \eqref{eq:meanfieldDicke} (black line).
For $N=100$ we find that even though the steady state agrees very well with the mean-field prediction, during dynamics, expectation values deviate significantly from our exact results. Additionally, a number of quantum trajectories localize on the opposing side of the sphere, such that the true $\rho$ relaxes to an uneven mixture of the two steady states predicted by mean-field. 
Going to larger $N$, we find for $N=500$ that even though the qualitative behaviour of the mean-field prediction is correct, there is still a quantitative difference in the dynamics close to the steady state. Individual quantum trajectories, however, follow the mean-field solution quite closely. 
This apparent contradiction is resolved by looking at the spin-Q function $Q(\theta, \varphi) = \bra{\theta, \varphi}\rho\ket{\theta, \varphi}$, which is shown at a fixed time $t_1 = 7.5/\omega_a$ in the lower plots (a2)-(c2) for the parameters corresponding to (a1)-(c1) respectively.  Here, $\ket{\theta, \phi}$ denotes a spin coherent state defined as $\ket{\theta, \phi} = \exp\left((\theta e^{i\phi} S^- -\theta e^{-i\phi}S^+)/2\right)\ket{j,j}$.
The Q-function shows how the individual trajectories are spread out along, but not perpendicular to the mean-field prediction \cite{unitary_Dicke_close_to_classical}, which suggests an effective approximation via truncated Wigner methods \cite{Polkovnikov2010Aug,Huber2022Jan,Mink2023Dec} is possible.
In (a2), (b2) the spin-state $\rho$ is thus highly non-Gaussian before the steady state is reached, leading to the breakdown of the mean-field approximation.
However, the spin state becomes more and more localized with increasing $N$ such that for $N=1000$ there is only a marginal difference between the exact dynamics and the mean-field solution.
For the balanced Dicke Model we are thus able to compute the entire range from fully quantum to effectively classical behaviour in a numerically exact way.
Finally, plots (d1), (d2) show the evolution for $N=1000$ starting from the normal state, which is the typical experimental initial condition. 
As this initial state obeys the symmetry $a\to -a$, $S^x,\, S^y \to -S^x,\,-S^y$ of the Hamiltonian \eqref{eq:Dicke_H} the evolution is restricted and mean-field can't be directly applied. The exact spin state $\rho_{sys}$ extends over a large part of the Hilbert space during the dynamics, so that density operator approaches are not feasible. Still, individual trajectories are localized, allowing for an efficient numerical representation in an adaptive basis \cite{PRA_placeholder}.

Having the full state at hand we can now turn to the question whether correlations and genuine quantum effects like entanglement play a role in the dynamics.
For that we focus on the atom-field covariance $C_{af}=\cavg{S^xa}-\cavg{S^x}\cavg{a}$, which is assumed to be zero for the mean-field equations \eqref{eq:meanfieldDicke}.
\begin{figure}
    \begin{minipage}{0.45\textwidth}
        \centering
             \includegraphics[width=\textwidth]{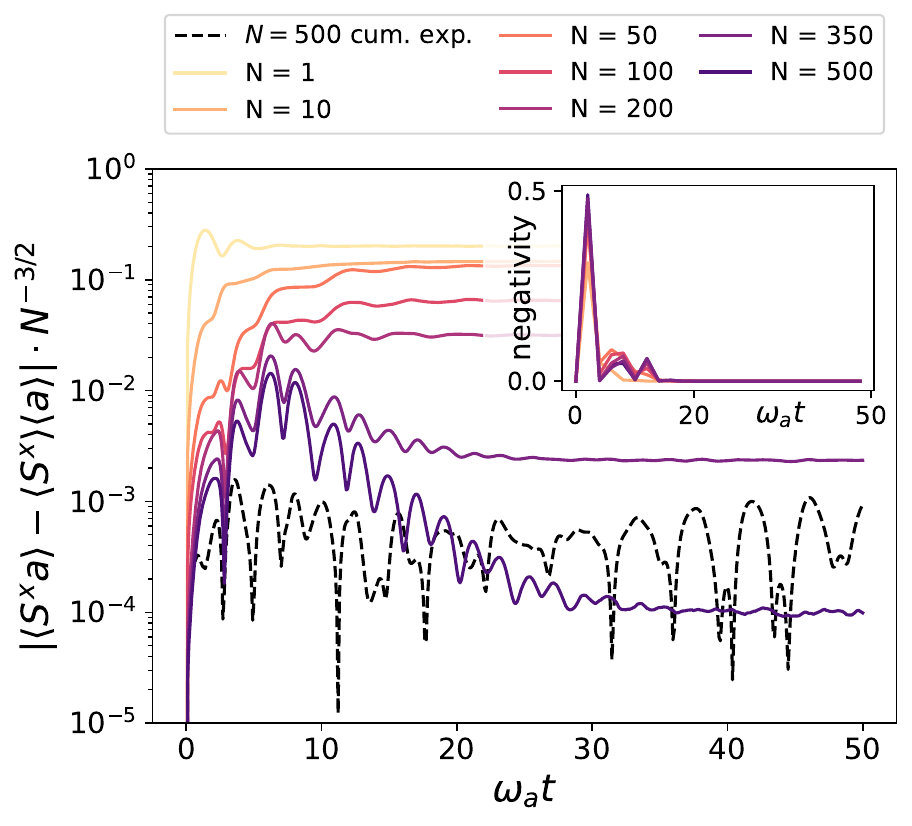}
    \caption{Time evolution of the covariance between the atoms and the field mode for different $N$ (solid) as well as the result of a second order cumulant expansion for $N=500$ (black dashed). For the inset we divide the $N$ spins in two groups of equal size and quantify the entanglement between the two groups with the help of the negativity of the partial transpose. This supports the view that the steady state correlations are of classical origin and true quantum correlations are only present during transient dynamics.}
    \label{fig:corrDynamicsDicke}
    \end{minipage}
\end{figure}
Figure \ref{fig:corrDynamicsDicke} shows the dynamics of $C_{af}/N^{3/2}$ for the initial product state $\ket{\Psi(0)} = \ket{\theta=\pi/4, \phi=\pi}\otimes\ket{0}$ and different numbers of atoms. 
Other parameters are identical to Fig.~\ref{fig:blochDynamics}. We compare our exact results to a second order cumulant expansion \cite{quantum_cumulants_jl}, which approximates the joint atom-field state as a Gaussian by neglecting cumulants of 3rd order or higher. 
Even though the cumulant expansion takes finite size effects of lowest order into account, it clearly fails to predict the covariance in this regime.
This can be traced back to fact, that while the state does become more localized for increasing $N$, it still maintains non-Gaussian features, as seen in Fig.~\ref{fig:blochDynamics}(a2) - (d2). 
In this parameter regime we thus find no significant advantage of using a cumulant expansion over the mean-field equations.\\
To investigate the occurrence of genuine quantum correlations we look at the entanglement between the spins, after dividing them in two groups of equal size. We quantify this by writing the state $\rho$ (describing a single spin-$N/2$) in a basis of two spins with length $N/4$ each and compute the negativity of the partial transpose \cite{negativity} of this two-spin representation. The result is shown in the inset of Fig.~\ref{fig:corrDynamicsDicke} and clearly shows how quantum correlations are only present in the transient dynamics before reaching the steady state. The steady state itself is very well described by the mean-field predictions, as can be seen from the rapidly decaying covariance in Fig.~\ref{fig:corrDynamicsDicke}.
\paragraph{Generalized Open Dicke Model.---}
We now turn to the more general case $\Delta g \neq 0$, where the atoms couple to both quadratures of the light field. 
It has been shown experimentally \cite{gen_Dicke_Singapur, genDicke_ETH} and theoretically \cite{gen_Dicke_semiclassic, gen_Dicke_phases_Zilberberg}, that the unbalanced Dicke Model has a much richer phase diagram, featuring limit cycles, and bistability regions in addition to the normal and superradiant phases.\\
In the following we will focus on the bistable region of the parameter space, which is expected to show spin squeezing \cite{gen_Dicke_finite_T} and recently gained attention 
in the study of quantum chaos \cite{gen_Dicke_coin_toss}. In the bistable phase both the normal state and the superradiant states are stable fixed points of the mean-field dynamics.
However, for finite $N$ the quantum state can tunnel between the different steady states predicted by mean-field and thus evolves towards a classical mixture of the normal and superradiant states, as shown in Fig.~\ref{fig:pdGenDicke}(a). 
\begin{figure*}[t]
    {
        \centering
        \begin{subfigure}[b]{0.35\textwidth}
             \centering
             \begin{overpic}[width=\linewidth]{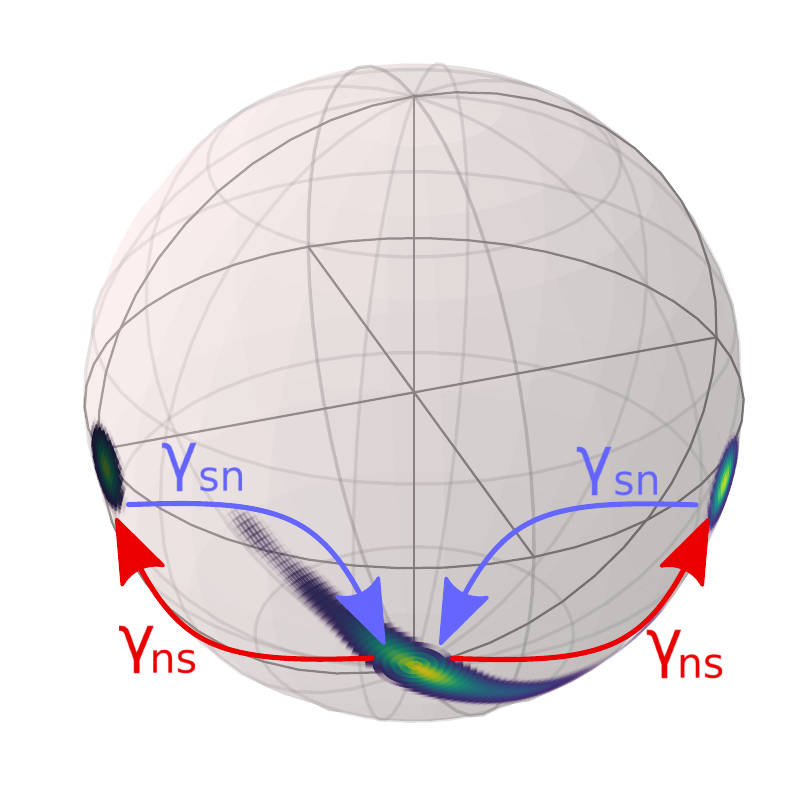} 
            \put(5,90){\textbf{(a)}} 
        \end{overpic}
        \end{subfigure}
        \begin{subfigure}[b]{0.35\textwidth}
        \centering
        \begin{overpic}[width=\linewidth]{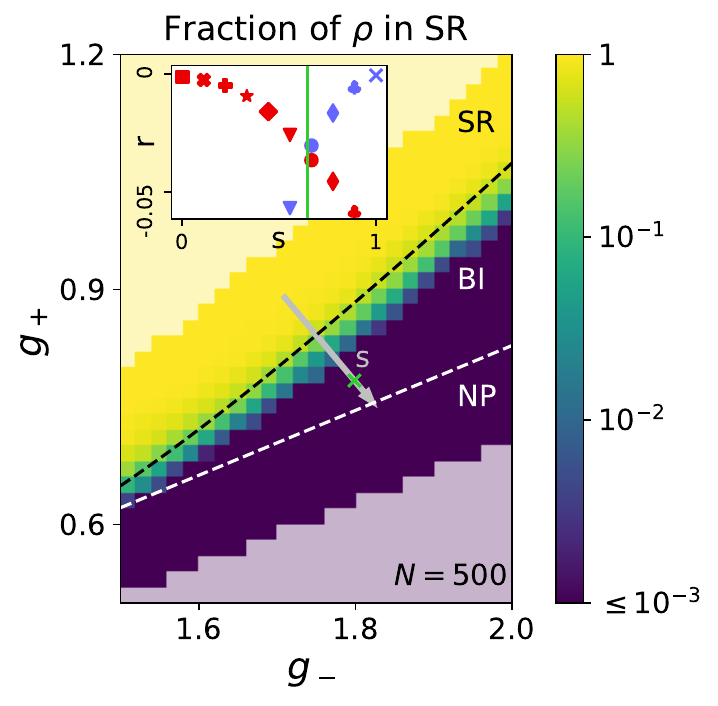} 
            \put(-3,90){\textbf{(b)}} 
        \end{overpic}
         \end{subfigure}
         \begin{subfigure}[b]{0.25\textwidth}
             \centering
        \begin{overpic}[width=\linewidth]{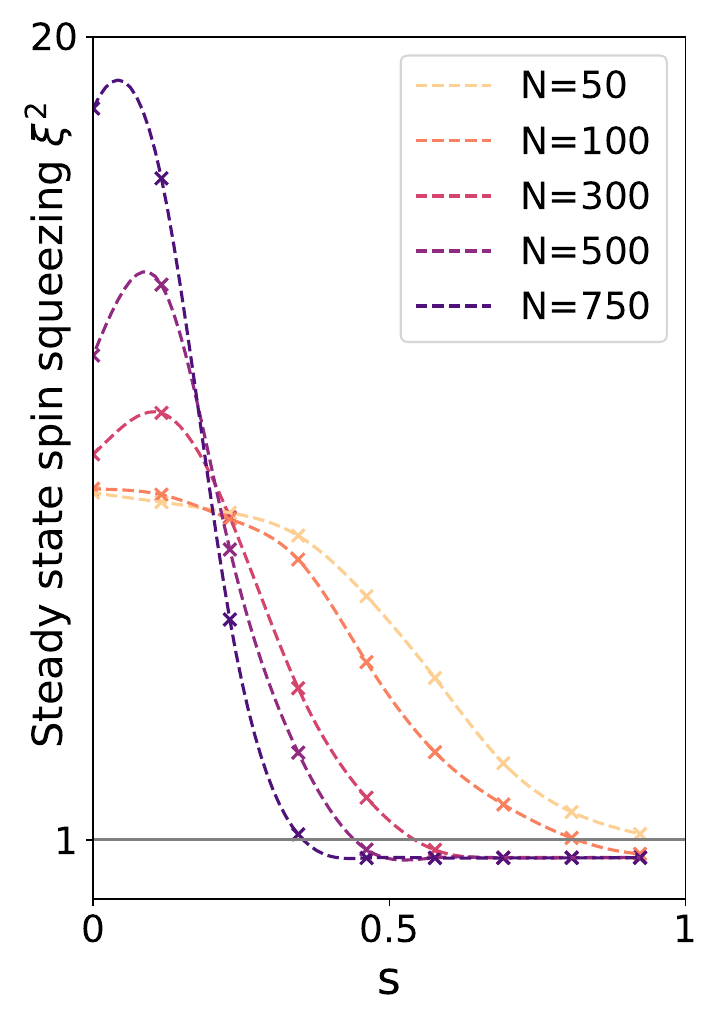} 
            \put(-8,90){\textbf{(c)}} 
        \end{overpic}
         \end{subfigure}
         \begin{subfigure}[b]{0.49\textwidth}
             \centering
        \begin{overpic}[width=\linewidth]{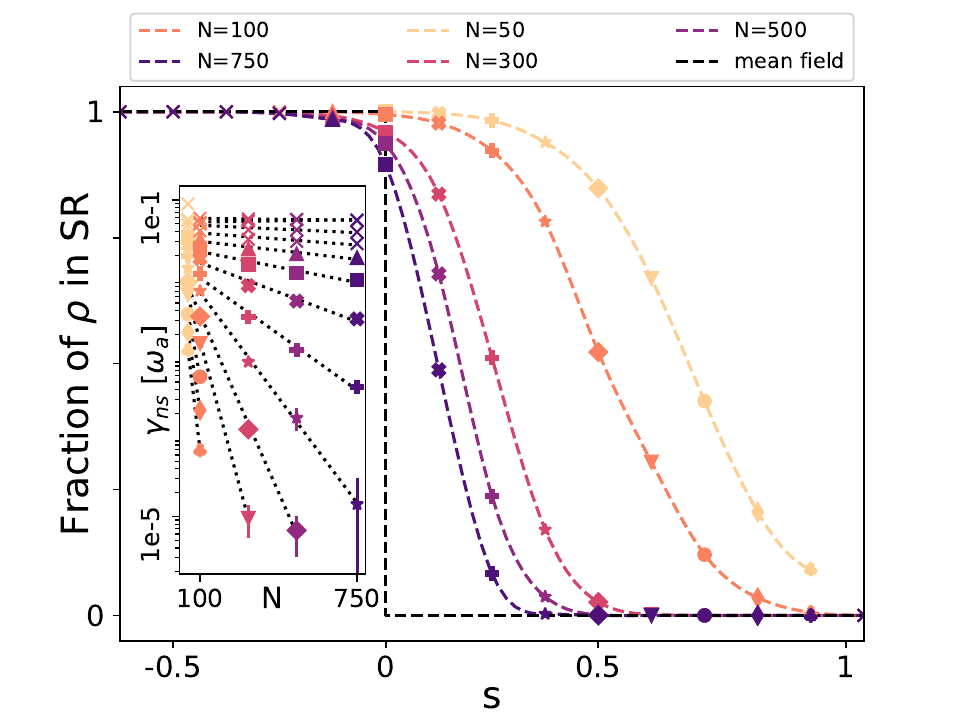} 
            \put(1,70){\textbf{(d)}} 
        \end{overpic}
        \end{subfigure}
         \begin{subfigure}[b]{0.49\textwidth}
             \centering
        \begin{overpic}[width=\linewidth]{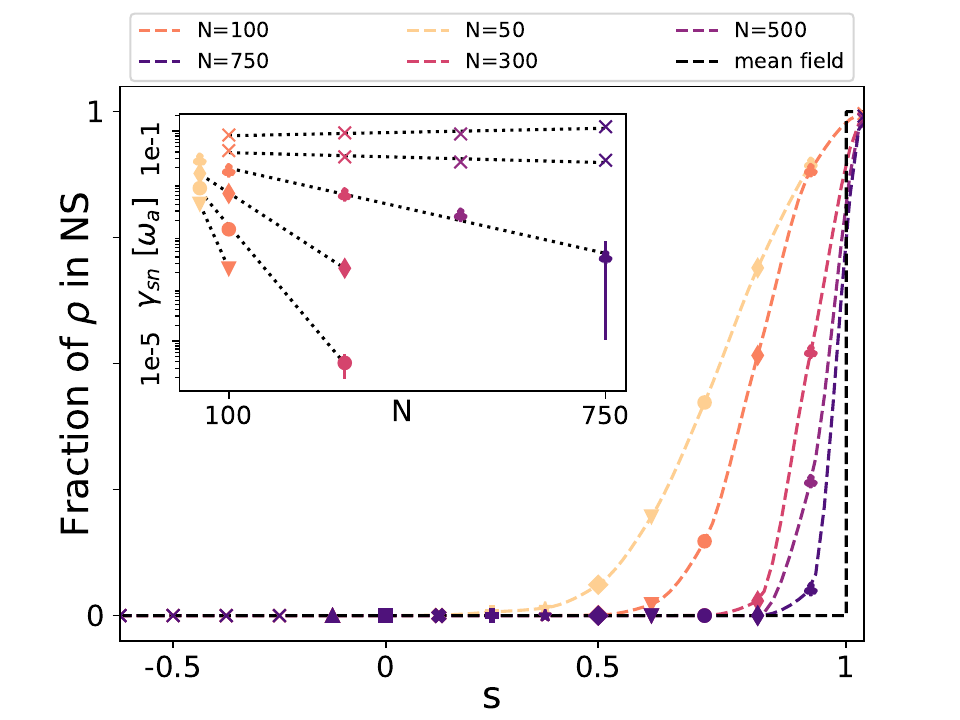} 
            \put(1,70){\textbf{(e)}} 
        \end{overpic}
         \end{subfigure}
    }
    \caption{Finite size effects on the bistable phase of the unbalanced Dicke model. (a) Example of a spin-Q function for $N=500$ in the bistable phase, showing the possibility of quantum tunneling. (b) Occupation of the superradiant state after starting in the normal state and evolving for $\omega_a t = 50$ for different parameters and $N=500$. Dashed lines show the mean-field boundaries of the superradiant (SR), bistable (BI) and normal (NP) phases. Inset shows a fit of the rate exponents $\gamma_i \propto \exp\left( r_i N \right)$ for $i\in \{\mathrm{ns},\mathrm{sn}\}$ and their intersection which marks the sharp transition in the quantum phase diagram and is shown as a green cross. (c)/(d) Occupation of the superradiant/normal state after starting in the normal/superradiant state and evolving for $\omega_a t= 150$ for different $N$ along a cut through the phase diagram. Color of the data points encodes the corresponding number of atoms and shape encodes the position along the cut. Colored dashed lines are obtained from a cubic spline fit and serve merely as a guide to the eye. Black dashed line corresponds to the mean-field prediction. Insets show the same data points as a plot with respect to $N$ together with the fit of the rate exponents $r_i$.  }
    \label{fig:pdGenDicke}
\end{figure*}
Starting the dynamics in the normal state the finite tunneling rate to the superradiant state $\gamma_{ns}$ leads to a finite occupation of the superradiant state after a given time, as shown in the finite size phase diagram in Fig.~\ref{fig:pdGenDicke}(b) for $N=500$ and $\omega_a t = 50$. 
There, the dashed black and white lines mark the mean-field boundaries of the respective normal, superradiant and bistable phases. The background color encodes the fraction of the state, which has tunneled to the superradiant state. 
The finite size effects are still clearly visible for $N=500$ and close to the superradiant phase quantum tunneling can be frequently observed even for larger atom numbers.
The dependence on the atom number is shown in Fig.~\ref{fig:pdGenDicke}(c), which shows the tunneling fraction on a cut through the phase diagram for different $N$. 
The location of the cut is marked as a grey arrow in Fig.~\ref{fig:pdGenDicke}(b) and parameterized by $s$, such that $s=0$ corresponds to the mean-field boundary to the superradiant phase and $s=1$ corresponds to the mean-field boundary to the normal phase. The colored dashed lines were obtained from a cubic spline interpolation of the data points and serve merely as a guide to the eye. Clearly, for this fixed time the finite size curves approach the sharp mean-field transition for $N\to\infty$.\\
In the following we are interested in the long time dynamics, to see if any quantum effects can survive in the steady state. As the state mainly evolves around the mean-field fixed points the long time dynamics is dictated by the tunneling rates $\gamma_{ns}$ and $\gamma_{sn}$ between the superradiant and the normal state.
In the Supplemental Material we show how for large $N$ and $t$ the population of the normal $p_n$ and superradiant state $p_s = 1 - p_n$ is given by

\begin{equation}\label{eq:rateeq}
\begin{split}
    p_n(t) =& \frac{\gamma_{sn}}{2\gamma_{ns}+\gamma_{sn}}+\frac{2\gamma_{ns}p_n(0) - \gamma_{sn}p_s(0)}{2\gamma_{ns}+\gamma_{sn}}e^{-(2\gamma_{ns} + \gamma_{sn})t}.
\end{split}
\end{equation}
Solving Eq.~\eqref{eq:rateeq} for the steady state we find a statistical mixture between the normal and superradiant state, where $p_n = 1/(1+2\gamma_{ns}/\gamma_{sn})$ and $p_s = 1/(1+0.5\gamma_{sn}/\gamma_{ns})$. \\
By fitting our numerical data to the solution of Eq.~\eqref{eq:rateeq} we can obtain the resulting tunneling rates as shown in the Supplemental Material.
The tunneling rates $\gamma_i$ have an exponential dependence on $N$ as $\gamma_i = A_i \exp{r_iN}$ $(r_i < 0)$, as is also shown in the insets of Fig.~\ref{fig:pdGenDicke}(c) and (d). 
From this we extrapolate the quantum phase diagram by taking the limit $N\to \infty$. 
This leads to an interesting consequence for the steady state occupations
\begin{equation}
    \begin{split}
        p_n &= \frac{1}{1+\frac{2A_{ns}}{A_{sn}}\exp{(r_{ns} - r_{sn})N}}\\
        &\overset{N\to\infty}{=}
        \begin{cases}
    0,& \text{if } r_{ns}>r_{sn}\\
    1,              & \text{if } r_{ns}<r_{sn}
\end{cases}
    \end{split}
\end{equation}
If we account for a finite tunneling probability the state thus undergoes a sharp transition from the normal to the superradiant phase at a position that is dictated by the tunneling rates and lies in between the two mean-field boundaries of the bistable phase ($s=0$ and $s=1$). We can estimate the exact position of the transition along the cut from our finite size data by fitting $r_{ns}$ and $r_{sn}$ as shown by the dashed lines in the insets of Fig.~\ref{fig:pdGenDicke}(c) and (d). The resulting exponents are plotted as an inset in Fig.~\ref{fig:pdGenDicke} and their intersection at $s_c \approx 0.6$, which marks the quantum phase transition, is shown by a green cross in the phase diagram. For $s$ above (below) $s_c$ the superradiant (normal) state is at large times exponentially suppressed in $N$, leading to a sharp transition in the thermodynamic limit. \\
The discrepancies between the mean-field and the genuine quantum phase diagram can be traced back to the order of taking limits. 
In the mean-field case one takes first the limit $N\to\infty$, thereby eliminating all quantum effects, and then takes the limit $t\to\infty$ by calculating the steady state. 
Here we take the limit $t\to\infty$ first for arbitrary finite $N$, by calculating the steady state of Eq.~\eqref{eq:rateeq}, and then consider the limit $N\to\infty$ of the finite size steady states. 
In fact, both limits are contained in Eq.~\eqref{eq:rateeq}, which gives the probabilities for all physical situations of large but finite $N,t$.\\
Although this breakdown of the bistable phase is intellectually fascinating, it is not observable in current experiments, where the state is probed after a finite time (much shorter than the inverse tunneling rate), similar to our simulation in Fig.~\ref{fig:pdGenDicke}(b).\\
To observe the breakdown long observation times combined with moderate atom numbers would be required.
However, while an observation of the sharp transition in the bistable phase is extremely challenging, it may be possible to find signatures of tunneling close to the boundaries of the mean-field-bistable phase. 
There, tunneling rates can become significant and one would thus expect a shift of the boundary depending on the measurement time.\\
Other traces of quantum behaviour can be found in the normal state, which exhibits strong squeezing in the bistable phase. We quantify this in Fig.~\ref{fig:pdGenDicke}(c).
We find that the spin-squeezing parameter $\xi^2$ \cite{Sorensen2001Jan,Wang2001Jun} stays below one even for large $N$, signaling entanglement between the spins in the steady state.

\paragraph{Conclusions.---}
We have applied a near-unitary (nuHOPS) extension of the hierarchy of pure states (HOPS)  \cite{HOPS,HOPS_Richard} to the (unbalanced) Dicke Model, which allows us to compute numerically exact dynamics from fully quantum ($N\approx1$) to effectively classical ($N=1000$) scenarios. 
The results for the balanced Dicke Model show clearly how the mean-field equations provide an excellent approximation for dynamics for $N\geq 1000$. 
However, contrary to what would be expected from looking at steady state properties alone, significant differences in the dynamics can arise if the number of spins is of the order of 100. 
In the bistable phase of the unbalanced Dicke Model we can find genuine quantum effects for even larger numbers of atoms and also in the steady state. 
These manifest themselves in a squeezing of the normal state and quantum tunneling in the vicinity of the mean-field boundaries of the bistable phase. 
In fact, these tunneling phenomena strongly influence the steady-state in the bistable phase for all $N$ - an effect that is completely absent in the mean-field phase diagram.
Surprisingly, it leads to the complete breakdown of the bistable phase, which is replaced by a sharp, discontinuous transition in the full quantum model. We note in passing that a truncated Wigner description fails to capture the correct tunneling rates. The question how to experimentally reach the required parameter regime as well as the exact details of this transition provide an exciting prospect for future studies.
In addition, our work opens up the possibility of numerically exact solutions even for cavity QED systems beyond the single collective spin paradigm, such as individual clumps of multiple atoms interacting via multimode cavities. 
Furthermore the ability to obtain exact solutions of a non-integrable many-body model out of equilibrium from the fully quantum up to the classical regime 
is a novelty that can give new insights into the quantum to classical transition, especially since the balanced as well as the unbalanced Dicke model are known to host chaotic phases.

\paragraph{Acknowledgments.---}
It is a pleasure to thank Valentin Link and Francesco Piazza for enlightening discussions in connection with this work. The authors gratefully acknowledge the computing time made available by the high-performance computer at the NHR Center of TU Dresden. This center is jointly supported by the Federal Ministry of Education and Research and the state governments participating in the NHR (www.nhr-verein.de/unsere-partner).

\bibliography{literature}
\clearpage
\onecolumngrid

\setcounter{equation}{0}
\renewcommand{\theequation}{S\arabic{equation}}
\renewcommand{\thesection}{SM\arabic{section}}
\renewcommand{\thefigure}{S\arabic{figure}}

\begin{center}
{\large \bf Supplemental Material}
\end{center}
\section{Spin squeezing}\label{app:spin_squeezing}
In order to quantify the amount of spin-squeezing we compute the spin-squeezing parameter
\begin{equation}
    \xi^2 = N\frac{(\Delta S^\perp)^2}{\cavg{S^{(1)}}^2 + \cavg{S^{(2)}}^2},
\end{equation}
where $S^\perp = \vec{{S}}\vec{e}_\perp$, $S^{(1)}=\vec{{S}}\vec{e}_{(1)}$, $S^{(2)}=\vec{{S}}\vec{e}_{(2)}$ with the three orthogonal unit vectors $\vec{e}_{\perp},\vec{e}_{(1)},\vec{e}_{(2)}$ and $\vec{\cavg{S}}\vec{e}_\perp = 0$. For each state the value of the spin-squeezing parameter is obtained from a minimization of $\xi^2$ over all possible vectors $\vec{e}_\perp$, which are perpendicular to the expectation value $\vec{\cavg{S}}$.

\section{Rate equation and numerical fit of the rates $\gamma_{ns}$, $\gamma_{sn}$}\label{sec:app_rate}
\begin{figure*}
    {
        \centering
        \begin{subfigure}[b]{0.49\textwidth}
             \centering
             \includegraphics[width=\textwidth]{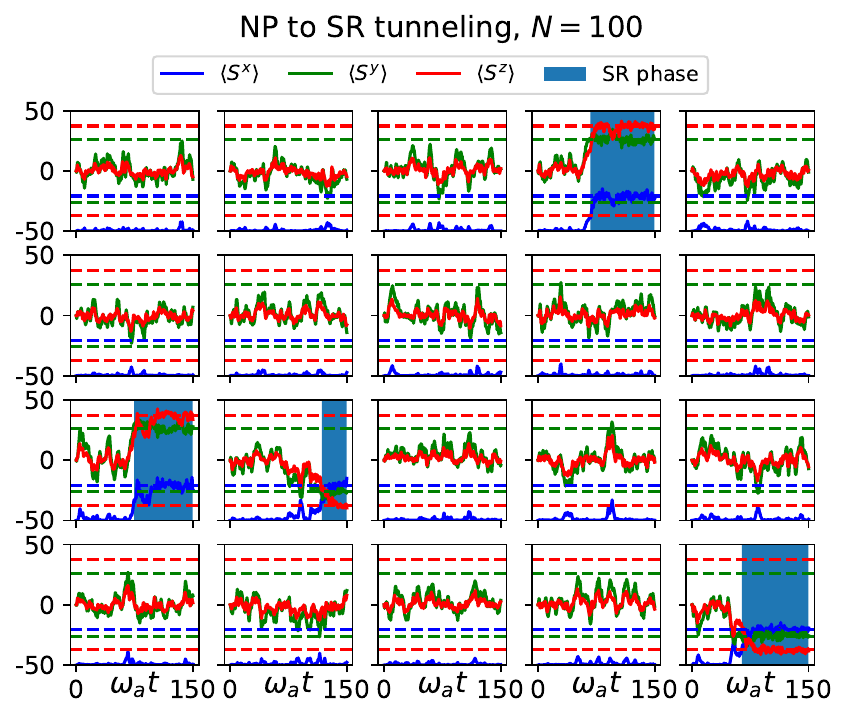}
         \end{subfigure}
         \begin{subfigure}[b]{0.49\textwidth}
             \centering
             \includegraphics[width=\textwidth]{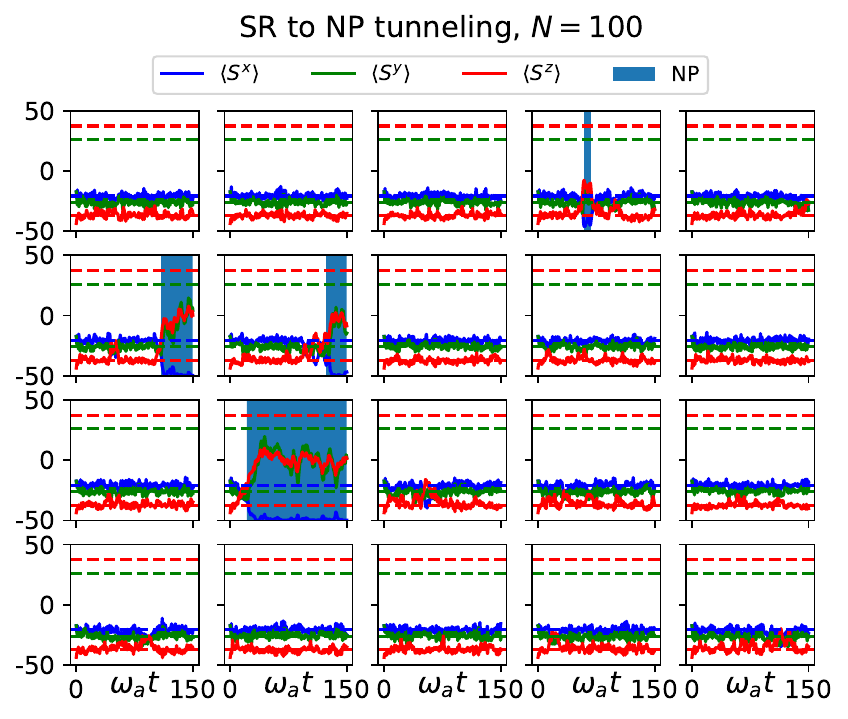}
         \end{subfigure}
         \begin{subfigure}[b]{0.98\textwidth}
             \centering
             \includegraphics[width=\textwidth]{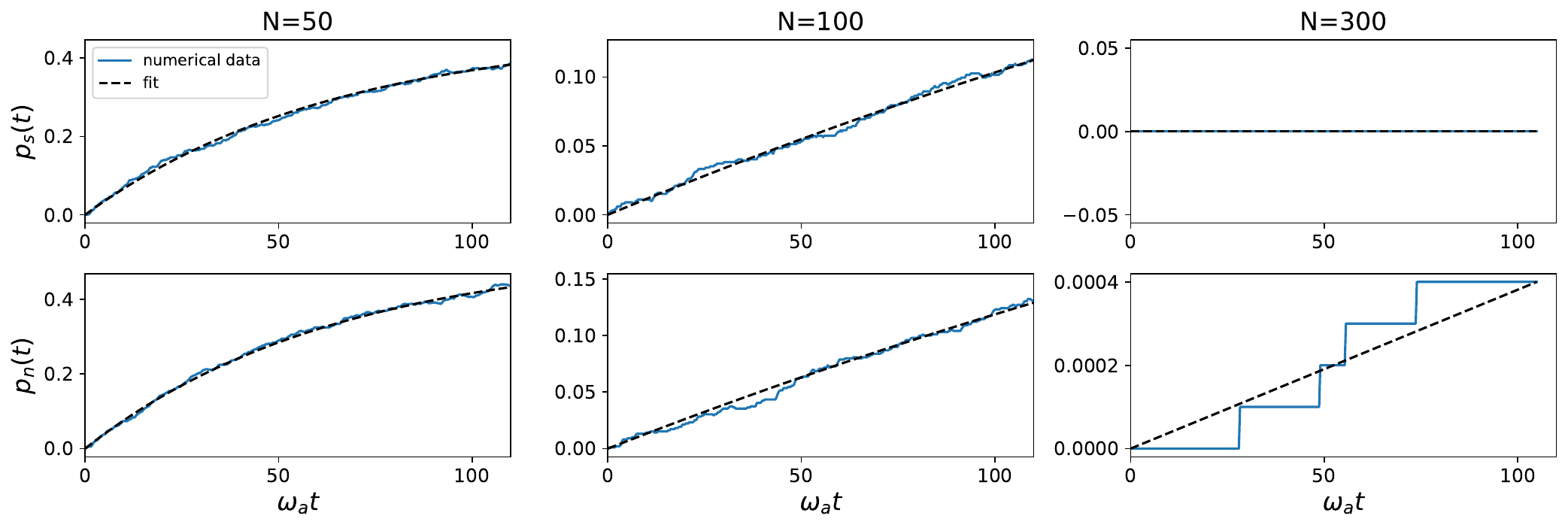}
         \end{subfigure}
    } 
    \caption{Example data for how the tunneling rates were obtained for the parameters $g_- = 1.8$, $g_+ = 0.782$ and $\omega_a = \omega_c = \kappa = 1$, corresponding to the data points at $s=0.69$ in Fig.~\ref{fig:pdGenDicke} (marked by circles, close to the crossing of the exponents). The upper left figure shows some individual trajectories obtained from a simulation with $N=100$ atoms starting in the normal state. The dashed lines mark the steady state expectation values in the superradiant phase. While most trajectories remain in the normal phase some tunnel into the superradiant phase marked by the blue background. The upper right figure shows trajectories starting in the superradiant state, with tunneling events to the normal phase marked in blue. The lower figure shows the probability to find the state in the initially unoccupied phase over time for different $N$ and both initial states. A fit to the solution of the rate equation \eqref{eq:app_rate_eq} shows very good agreement and is used to extract the tunneling rates from the numerical data. Note that for this combination of $g_+$, $g_-$ no tunneling to the superradiant state was observed for $N=300$.}
    \label{fig:app_rate_fit}
\end{figure*}

In the following we provide detailed information on how we obtain the tunneling rates $\gamma_{sn}$, $\gamma_{ns}$ from the numerical data. 
Based on the assumption that the tunneling rates are much smaller than the cavity decay rate $\kappa$, we propose a Markovian rate equation for the long time dynamics of the system. As shown in Fig.~3 a) there are three meta stable states in the bistable phases. We denote the probability of finding the state in the superradiant state with $\cavg{S^x} > 0$ ($\cavg{S^x}<0$) with $p_{s+}$ ($p_{s-}$) and the probability of finding the state in the normal phase with $p_n$. Since the two superradiant steady states are related by the symmetry $S^x\to-S^x$, $S^y \to -S^y$, $a\to-a$ their tunneling rates to and from the normal state must be identical. The rate equation then takes the form
\begin{equation}\label{eq:app_rate_full}
    \begin{split}
        \dot{p}_n =& -2\gamma_{ns}p_n + \gamma_{sn}(p_{s+} + p_{s-}),\\
        \dot{p}_{s+} =& -(\gamma_{sn} + \gamma_{ss})p_{s+} + \gamma_{ss}p_{s-} + \gamma_{ns}p_n,\\
        \dot{p}_{s-} =& -(\gamma_{sn} + \gamma_{ss})p_{s-} + \gamma_{ss}p_{s+} + \gamma_{ns}p_n,
    \end{split}
\end{equation}
where $\gamma_{ss}$ denotes the tunneling rate between the two superradiant steady states. For the probability of being in any of the two superradiant states $p_s = p_{s+} + p_{s-}$ we thus find
\begin{equation}\label{eq:app_rate_eq}
    \begin{split}
        \dot{p}_n =& -2\gamma_{ns}p_n + \gamma_{sn}p_s,\\
        \dot{p}_{s} =& -\gamma_{sn}p_{s} + 2\gamma_{ns}p_n.
    \end{split}
\end{equation}
For an initial state in the normal phase we find the solution
\begin{equation}\label{eq:app_n-s}
    p_s(t) = \frac{2\gamma_{ns}}{2\gamma_{ns}+\gamma_{sn}}\left(1-e^{-(2\gamma_{ns} + \gamma_{sn})t}\right),
\end{equation}
and $p_n(t) = 1-p_s(t)$. Conversely if the initial state is in the superradiant phase we find
\begin{equation}\label{eq:app_s-n}
    p_n(t) = \frac{\gamma_{sn}}{2\gamma_{ns}+\gamma_{sn}}\left(1-e^{-(2\gamma_{ns} + \gamma_{sn})t}\right).
\end{equation}
In order to obtain the tunneling rates we now perform numerical simulations starting in the normal as well as in the superradiant state for different parameters. As an example, a few trajectories obtained from these simulations are shown in Fig.~\ref{fig:app_rate_fit}. 
These trajectories were calculated with the parameters $N=100$, $g_- = 1.8$, $g_+ = 0.782$ and $\omega_a = \omega_c = \kappa = 1$, which corresponds to the data points at $s=0.69$ in Fig.~\ref{fig:pdGenDicke} (marked by circles, close to the crossing of the exponents). 
In the upper left figure we start the simulation in the ground state and check for tunneling events to the superradiant phase (marked by the blue background). 
We detect the tunneling events by comparing a moving average of the expectation values $\cavg{S^i}$ to the expected steady state values indicated by the dashed lines. 
We proceed analogously for the superradiant initial state (trajectories shown in the upper right figure), where we check for tunneling events to the normal phase.\\
After we classified the phase at each time step for each trajectory into being either normal or superradiant we can take the average to obtain a stochastic estimate of the probabilities $p_s(t)$ (for trajectories starting in the normal state) and $p_n(t)$ (for trajectories starting in the superradiant state), which are shown in the lower figure. 
We then fit the analytical solutions \eqref{eq:app_n-s} and \eqref{eq:app_s-n} of the rate equation to the two numerical curves obtained for each set of parameters to obtain the tunneling rates $\gamma_{ns}$ and $\gamma_{sn}$. 
We find that the fitted analytical solutions match very well to the numerical data, confirming that the rate equation \eqref{eq:app_rate_eq} provides an adequate description of the long-time dynamics. 
The obtained rates are shown in the insets of Fig.~3 (d) ($\gamma_{ns}$) and Fig.~3 (e) ($\gamma_{sn}$) of the main paper.
\end{document}